\documentclass[twocolumn]{jpsj2}
\def\runauthor{B. Lee Roberts}

\title{
Muon Physics: A Pillar of the Standard Model
}

\author{
B. Lee Roberts\thanks{E-mail address: roberts@bu.edu}
}

\inst{
Department of Physics\\
Boston University\\
Boston, MA 02215, USA
}

\recdate{\today}
\abst{%
Since its discovery 
in the 1930s, the muon has played an important role in 
our quest to understand the sub-atomic theory of matter.  The muon was
the first second-generation standard-model particle to be discovered,
and its decay has provided information on the $(Vector\ -\ Axial\ Vector)$
structure of the weak interaction, 
 the strength of the weak interaction, $G_F$, and
the conservation of lepton number (flavor) in muon
decay. The muon's anomalous magnetic
moment has played an important role in restricting theories of physics
beyond the standard standard model, where at present there is a 3.4~$\sigma$
difference between the experiment and standard-model theory.
Its capture on the atomic nucleus
has provided valuable information on the modification of the weak 
current by the strong interaction which is complementary to that obtained
from nuclear $\beta$ decay.
}

\kword{%
muon, weak decay, muon capture, magnetic moment, lepton flavor violation
}

\begin{document}
\maketitle

\section{Introduction}

The muon was first observed in a Wilson cloud chamber by Kunze\cite{kunze}
in 1933, where it was reported to be ``a particle of 
uncertain nature.''  In  1936  Anderson and 
Neddermeyer\cite{Ned} reported the presence of ``particles less massive
than protons but more penetrating than electrons'' in
cosmic rays, which was
confirmed  in 1937 by 
Street and Stevenson\cite{Street}, Nishina, Tekeuchi and 
Ichimiya\cite{Nishina}, and by Crussard and Leprince-Ringuet\cite{Crussard}.
The Yukawa theory of the nuclear force had predicted such a particle,
but this ``mesotron'' as it was called, interacted too weakly with 
matter to be the carrier of the strong force.  
Today we understand that the muon is a second generation lepton, with
a mass about 207 times the electron's.
Like the electron, the muon obeys quantum electrodynamics, and
can interact with other particles  through the electromagnetic
and weak forces.  Unlike the 
electron which appears to be stable, the muon decays through the weak force.  

The muon lifetime
of $2.2\ \mu$s permits one to make precision measurements of its
properties, and to use it as a tool to study the semileptonic
weak interaction,  nuclear properties,  as well as
magnetic properties of condensed matter systems. 
The high precision to which the muonium ($\mu^+ e^-$ atom)
hyperfine structure can be measured and calculated
makes it a significant input parameter in the determination of
fundamental constants\cite{codata}.
In this review, I will focus
on the role of the muon in particle physics.

A beam of negative muons can be brought to rest in matter, where
hydrogen-like atoms are formed, with a nuclear charge of $Z$.
The Bohr radius for a hydrogen-like
atom is inversely proportional to the orbiting particle's mass
($r_n = [n^2 \hbar c] /[mc^2 Z \alpha$] ), so that for
the lowest quantum numbers of high-$Z$
muonic atoms, the muon is well inside of the atomic electron cloud,
with the Bohr radius of the $1S$ atomic state
well inside the nucleus.  The $2P \rightarrow 1S$ x-ray energies
are shifted because of the modification of the Coulomb potential
inside the nucleus,
 and these x rays  have provided information on nuclear
root-mean-square charge radii.  
The Lamb shift in muonic hydrogen, $\Delta E_{2P-2S}$,
which is being measured at
at the Paul Scherrer Insitut (PSI),
is given 
by\cite{protonrad}
 $\left\{ 209.974(6) - 5.226 R^2_p + 0.036 R^3_p\right\}$~meV,
where $R_p$ is the proton rms charge radius.
This experiment
should provide a precise measurement of  $R_p$.  The weak nuclear 
 capture, called ordinary muon capture (OMC), 
of the muon on the atomic nucleus following the cascade
to the $1S$ ground state,
$\mu^- + _Z{\mathcal N} \rightarrow _{Z-1}{\mathcal N} + \nu_\mu$,
is the analog to the weak capture of a $K$-shell electron by the nucleus,
and provides information on the
modification of the weak interaction by the hadronic matter.

The muon mass of $\sim 106$~MeV
restricts the muon to decay into the electron, neutrinos, and photons.
Thus muon decay is a  purely leptonic process, and the dominant decay mode is
$\mu^- \rightarrow e^- +\nu_\mu + \bar \nu_e.$
This three-body decay tells us that the individual 
lepton number, electron and muon, is conserved separately, and that
the two flavors (kinds) of neutrinos are distinct 
particles\cite{mu-neut}. Here  the $\mu^-$ and $e^-$ are ``particles''
 and the  $\mu^+$ and $e^+$ are
the antiparticles.  In the 1950s, it became possible to make
pions, and thus  muons, in the laboratory.
The energetically favorable
decay $\mu^+ \rightarrow e \gamma$ was searched for and not 
found\cite{stein1} to a relative branching ratio of
$<2 \times 10^{-5}$.  Also searched for was the neutrinoless 
capture of a $\mu^-$ on an atomic nucleus,
 $\mu^- + {\mathcal N} \rightarrow e^- + {\mathcal N}$, which was not found
at the level of $\sim 5 \times 10^{-4}$.
Such processes are said to ``violate lepton flavor,'' and continue to be
the object of present and planned studies reaching to sensitivities 
of $10^{-14}$ and $10^{-16}$, respectively.

The muon, like the electron, is a spin $1/2$ lepton, with a
magnetic moment given by
\begin{equation}
 \vec \mu_s = g_s ( \frac {q} { 2m} ) \vec s;
\qquad \mu = (1 + a)\frac{q \hbar}{ 2m}; \qquad a 
\equiv \frac{ (g_s -2)}{ 2};
\end{equation}
where the muon charge $q= \pm e$,
and $g_s$, the Land\'e $g$-factor is slightly greater than the 
Dirac value of 2. The middle equation above is useful from a theoretical
point of view, as it separates the magnetic moment into two pieces:
the Dirac moment which is unity in units of
the appropriate magneton, $ {e \hbar / 2m}$, and is predicted by the Dirac
equation; and
the anomalous (Pauli) moment, where the dimensionless quantity
$a$ is referred to as  {\it the anomaly}.  The muon anomaly, like the
electron's, arises from radiative corrections that are discussed 
below.

When the muon was discovered, it was an unexpected surprise.
Looking at this from our 21st century perspective, it is easy to forget
how we reached what is now called the ``standard model'' of subatomic
physics, which incorporates three generations
of leptons, $e$, $\mu$ and $\tau$ and their neutrinos;
three generations of quarks; the electro-weak gauge bosons, $\gamma$
$W$ and $Z$; and the gluons that carry the strong force.
When this author joined the field as a graduate student in the mid
1960s, none of this was clear. Quarks were viewed by many as
a mathematical device, not as constituent particles. Even after
quarks were inferred from deep inelastic electron scattering off the proton, 
we only knew of the existence of
 three of them.  While the $V-A$ structure of the
weak interaction was first inferred from nuclear $\beta$ decay, the study of 
muon decay has provided a useful laboratory in which to study the 
purely leptonic weak interaction, to search for physics beyond the
standard model, such as additional terms in the interaction
besides the standard-model
 $V-A$ structure,
 as well as looking for standard model forbidden decays like
$\mu \rightarrow e \gamma$.
 For many years, the experimental value of
the muon's anomalous magnetic moment
has served to constrain physics beyond 
the standard model, and  continues that
role today.

\section{Muon Decay and $G_F$}

The muon decay $\mu^- \rightarrow e^- \nu_\mu  \bar \nu_e$ is purely leptonic.
Since $m_\mu << M_W$, muon decay can be described by a local
four-fermion (contact) interaction. While
nonrenormalizeable, at low energies
 it provides an excellent approximation to the 
full electroweak theory. The weak Lagrangian is written as a current-current
interaction,
where the leptonic current is of the ($V-A$) 
form, $ \bar u \gamma_\lambda ( 1 - \gamma_5)u$.

Michel\cite{michel} first wrote down a parameterization of muon decay,
defining five parameters, $\rho$, $\eta$, $\xi$, $\delta$ 
and $h$, which are combinations of the different possible couplings allowed
by Lorentz invariance in muon decay.  
The standard model has clear predictions for these parameters
and they have been measured repeatedly over the intervening
years to search for physics beyond the standard model. 
 This tradition continues today, with
 the TWIST experiment at TRIUMF, which is mid-way through a program to 
improve on the precision of the Michel parameters by an order of 
magnitude\cite{TWIST}.  While there are some scenarios in which new physics
would conspire to leave the Michel parameters at their standard model 
value\cite{cernwg},
a variance from the standard-model values would be a clear sign of new
physics at work.

\begin{figure}[t]
\begin{center}
\includegraphics[width=0.4\textwidth]{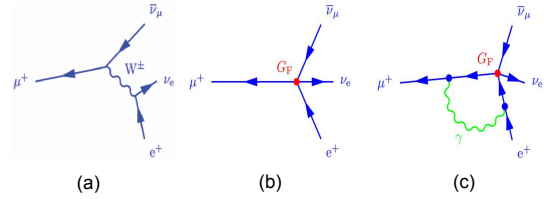}
\end{center}
\caption{Muon decay in (a) the standard model; (b) the Fermi theory;
(c) radiative corrections to the Fermi theory.}
\label{fg:mudecay}
\end{figure}

The muon lifetime, see Fig.~\ref{fg:mudecay} is directly related 
to the strength of the weak interaction, which in Fermi theory is
described by the constant $G_F$.
The standard-model electroweak
gauge coupling $g$ is related to $G_F$ by\cite{vR}
\begin{equation}
\frac{G_F}{ \sqrt 2} = \frac{g^2}{ 8 M_W^2} (1 + \Delta r)
\end{equation}
where $\Delta r$ represents the weak boson mediated tree-level
process
and its radiative corrections\cite{awramik}.
In the standard model, the Fermi 
constant is related to the vacuum expectation value of the 
Higgs field by $ G_F ={1}/({\sqrt{2}}~ v^2)$.

While the Fermi theory is nonrenormalizeable, the QED radiative
corrections are finite to first order in $G_F$, and to all orders 
in the fine-structure constant $\alpha$.  This gives the
relationship\cite{vR} between $G_F$ and the muon lifetime, $\tau_\mu$,
\begin{equation}
\frac{1} {\tau_\mu} = \frac{G_F^2 m\mu^5} { 192 \pi^3} ( 1 + \Delta q)
\end{equation}
where $\Delta q$ is the sum of phase space, and QED and
hadronic radiative corrections.
More properly one should write $G_\mu$ since new physics contributions
could make $G$ different for the three leptons.\cite{marcianoGF}

The MuLan experiment at PSI has recently reported 
a new measurement of the muon lifetime $2.197\,013(21)(11)\ \mu{\rm s}$
($\pm 11$ parts per million (ppm))\cite{Chitwood07}, 
to be compared with the previous world average
$2.197\,03(4)\ \mu{\rm s}$ (19~ppm)\cite{PDG}.  The new
world average muon lifetime of $2.197\,019(21)\ \mu{\rm s}$ gives,
assuming only standard-model physics in muon decay,
$G_F = 1.166\,371(6) \times 10^{-5}\ {\rm GeV}^2$ (5~ppm).
This experiment should eventually reach a precision of 1~ppm
on $\tau_\mu$.

\section{Nuclear Muon Capture}

The weak capture of a muon on a proton
 has much in common with nuclear
$\beta$ decay.  As for other low-energy weak processes, the interaction can 
be described as a current-current interaction with the 
$(V-A)$ leptonic current
given by $\bar u_{\nu_\mu} \gamma^\lambda ( 1 - \gamma_5) u_\mu$.
Because  the strong interaction can induce additional 
couplings\cite{Gorringe04}, 
the hadronic current is more complicated. The most
general form of the vector current allowed by Lorentz invariance 
is\cite{convention}
\begin{equation}
\bar u_n(p')\left [
g_V(q^2)\gamma^{\lambda} 
+ i \frac{ g_{M}(q^2)}{ 2 m_N}\sigma^{\lambda \nu}q_{\nu}
+ \frac{g_S(q^2)}{ m_\mu} q^{\lambda} \right]u_p(p).
\label{eq:hadvec}
\end{equation}
The corresponding form of the axial-vector current is
\begin{equation}
 \bar u_n (p')\left[
- g_A(q^2)\gamma^{\lambda}\gamma_5 
- \frac{ g_P(q^2)}{ m_\mu}\gamma_5 q^{\lambda}
-i \frac{g_T(q^2)}{ 2m_N}  \sigma^{\lambda \nu}q_{\nu}\gamma_5  
\right]u_p(p) ,
\label{eq:hadaxv}
\end{equation}
where $m_\mu$ and $m_N$ are the muon and nucleon masses respectively;
 the $g(q^2)$ are the induced form factors:
vector, weak magnetism, scalar, axial-vector, pseudoscalar and tensor.
The scalar and tensor terms are called ``second class currents'' because
of their transformation properties under $G$-parity, and in the standard
model are expected to be quite small\cite{Gorringe04}. It is traditional
to set these second-class currents equal to zero.

Nuclear $\beta$ decay is sensitive to the vector, axial-vector and 
weak-magnetism form factors, but in muon capture the
the capture rate has a measurable contribution from 
the  induced pseudoscalar interaction, the least well known
of the weak nucleon form factors.
Radiative muon capture (RMC), $\mu^- + p \rightarrow n + \gamma + \nu_\mu$,
should in principle be more sensitive
to the induced pseudoscalar coupling than OMC, since with the three-body final
state, $q^2$ can get closer to the pion pole than is possible in 
ordinary muon capture, which was
 pointed out many years ago\cite{manacher,opat}. The interested reader
is referred to the review by Gorringe and Fearing for further 
discussion\cite{Gorringe04}.
  
In the past, current algebra and the Goldberger-Trieman
relation expressed  $g_P$ in terms of $g_A$.  With the development of
quantum chromodynamics (QCD), and chiral perturbation theory, new interest
has developed in the value of $g_P$\cite{Gorringe04}. The presently 
accepted theory value is 
$g_P(q^2 = -0.88m^2_\mu)= 8.26 \pm 0.23$\cite{Gorringe04}.

The experimental history is rather interesting.  For many
years  the radiative
capture reaction $\mu^- + p \rightarrow n + \gamma + \nu_\mu$ was thought 
to be the ``golden'' channel to study.   However, this experiment
is extremely difficult, and it was first observed experimentally 
in the 1990s\cite{jonkmans,wright}.  To achieve an adequate
muon stopping rate, it was necessary to stop the muons a liquid hydrogen
target.  The value obtained for $g_p$ by this 
experiment was in disagreement with both the 
ordinary muon capture experiment in liquid hydrogen, and with the theoretical
expectation.

The muon chemistry in
hydrogen, especially liquid, is rather complicated, and may be
the source of these discrepancies.  The incident
$\mu^-$ first forms an atom with a proton in the hydrogen target, but then
quickly picks up a second proton to form a  
$p\mu p$ molecule with the protons in the ortho ($J=1$)
state. The ortho state of this molecule can decay to the 
 $J=0$ (para) state.  The ortho and para proton states have different
admixtures of the muon-proton spin: ortho has 3/4 singlet and 1/4 triplet
$\mu-p$ and para has 1/4 singlet and 3/4 triplet.  Since the muon
capture rate is 40 times greater from the singlet $\mu^-p$ state than from
the triplet, it is essential that $\Lambda_{OP}$, 
the ortho to para  transition rate  in the $p\mu p$ molecule, is known 
in order to extract $g_p$ from the measured capture rate in liquid
hydrogen\cite{Gorringe04}. Even after
a recent measurement\cite{clark} of the transition 
rate $\Lambda_{OP}$, the difficulty in 
accommodating previous results on ordinary and radiative
muon capture results in hydrogen continues.
The complications of muon chemistry in liquid hydrogen can be avoided
by using a 10 bar ultra-pure hydrogen target, which has a density 1.16\%
that of liquid hydrogen. At this lower density, the sensitivity
to $\Lambda_{OP}$ is greatly reduced. It is this approach that the
recent MuCap experiment at PSI has used.

The MuCap experiment stops $\mu^-$ in a  gaseous hydrogen target that
functions as a time projection chamber (TPC), making it possible to
 determine where the muon
stops in the target.  A comparison of the
 $\mu^-$ lifetime in this protonium target to the
the free $\mu^+$ lifetime, gives the capture rate and 
determines $g_p$.  The MuCap experiment  has 
recently reported a first result\cite{mucap}, 
$g_p(q^2=-0.88m^2_\mu) = 7.3 \pm 1.1$, 
consistent with the expectation from chiral perturbation theory.
They have a factor of four more data which are being analyzed.
While it is not clear what is wrong with the previous ordinary muon capture
and radiative capture experiments, the MuCap result seems to indicate that
a modern experiment, with a gaseous target and information from the
TPC, has settled the long-standing discrepancy.

\section{The Magnetic and Electric Dipole Moments}

The electric and magnetic dipole 
moments have been an integral part of relativistic electron
(lepton) theory since Dirac's famous 1928
paper, where he pointed out
that an electron in external electric and magnetic fields has 
``the two extra terms 
\begin{equation}
\frac{e h }{ c}\left({\mathbf \sigma} , {\mathbf H} \right) 
+ i \frac{e h}{ c}\rho_1 \left( {\mathbf \sigma} , {\mathbf E} \right) ,
\label{eq:dirac-dpm}
\end{equation}
\dots when divided by the factor $2m$ can be regarded as the
additional potential energy of the electron due to its new
degree of freedom\cite{Dirac}.''
 These terms represent the magnetic
dipole (Dirac) moment and electric dipole moment interactions with
the external magnetic and electric fields. 

In modern notation, the magnetic dipole moment (MDM) interaction becomes
\begin{equation}
\bar u_{\mu}\left[ eF_1(q^2)\gamma_{\beta} +
\frac{i e}{ 2m_{\mu}}F_2(q^2)\sigma_{\beta \delta}q^{\delta}\right] u_{\mu}
\end{equation}
where
$F_1(0) = 1,$ and $F_2(0) = a_{\mu}$. 
The electric dipole moment (EDM) interaction is
\begin{equation}\bar u_{\mu}
\left[\frac {i e}{ 2 m_{\mu}} F_2(q^2) - F_3(q^2)\gamma_5
\right]\sigma_{\beta \delta}
q^{\nu}u_{\mu}
\end{equation}
where $F_2(0) = a_{\mu}$, $ F_3(0) = d_{\mu}$, with
\begin{equation}
d_{\mu} = \left( \frac{\eta} {2} \right) \left(\frac {e \hbar }{ 2 mc}\right) 
\simeq \eta \times 4.7\times 10^{-14}\ e\,{\rm cm}.
\label{eq:eta}
\end{equation}
(This $\eta$, which is the EDM analogy to $g$ for the MDM,
 should not be confused with the Michel parameter $\eta$.)

The existence of an EDM implies that both {\sl P} and {\sl T} are 
violated\cite{ramsey,landau}.  This can be seen by considering the
 non-relativistic Hamiltonian for a spin one-half
 particle in the presence of
both an electric and magnetic field:
${\mathcal H} = - \vec \mu \cdot \vec B  - \vec d \cdot \vec E$.
The transformation properties of $\vec E$, $\vec B$, $\vec \mu$ and $\vec d$
are given in the Table \ref{tb:tranprop}, and we see that while
$\vec \mu \cdot \vec B$ is even under all three,
$\vec d \cdot \vec E$ is odd under both {\sl P} and
{\sl T}. While parity violation has been observed in many weak processes,
direct {\sl T} violation has only been observed in the neutral 
kaon system\cite{CPLEAR}.
In the context of {\sl CPT} symmetry, an EDM implies {\sl CP} 
violation, which is allowed by the standard model for decays
 in the neutral kaon and $B$-meson 
sectors.

\begin{table}[htb]
\caption{Transformation properties of the magnetic and electric fields and
dipole moments.}
\label{tb:tranprop}
\begin{tabular}{cccc} \hline
      & {$\vec E$ } &{$\vec B$ }  &{$\vec \mu$ or  $\vec d$} \\
\hline
{\sl P} & - & + & + \\
{\sl C} & - & - & - \\
{\sl T} & + & - & - \\
\hline
\end{tabular}
\end{table}

Observation of a non-zero electron or muon EDM would be a clear signal for
new physics.  To date no permanent EDM has been
observed for the electron, the neutron, or an atomic nucleus, with the 
experimental limits given in Table~\ref{tb:edm}.
It is interesting to note that in his original paper\cite{Dirac}
Dirac stated ``The electric moment, being a pure imaginary, we should
not expect to appear in the model.  It is doubtful whether the electric moment
 has any physical meaning, since the Hamiltonian 
\dots that we started from is real, and the imaginary part only appeared
when we multiplied it up in an artificial way in order to make it resemble
the Hamiltonian of previous theories.''  Even in the 4th edition of his
quantum mechanics book from 1958, well after the suggestion of Purcell
and Ramsey\cite{purcell} that one should search for a permanent EDM,
Dirac held fast to this point of view.

While {\sl CP} violation is widely invoked to explain the baryon-antibaryon
asymmetry of the universe, the {\sl CP} violation observed to date in
the neutral kaon, and in the $B$ meson sectors is too small to explain it.
  This
{\sl CP} deficit has motived a broad program of searches for EDMs in
a range of systems.  Many extensions to the standard model, such as
supersymmetry, do not forbid new sources of {\sl CP}-violation, and
the failure to observe it has placed severe restrictions on many 
models.

\begin{table}[h!]
\caption{Measured limits on electric dipole moments, and their standard
model values}
\label{tb:edm}
\centering
\begin{tabular}{ccc} \hline
   { Particle}  &{ Present EDM} & { Standard Model}  \\
                 & Limit ($e$~cm)           & Value ($e$~cm) \\
\hline
$n$ & {$2.9 \times 10^{-26}$ } (90\%CL)\cite{nedm}  & {$10^{-31}$ }  \\
\hline
 $e^-$  & {$\sim 1.6 \times 10^{-27 }$} (90\%CL)\cite{eedm} & {$10^{-38}$ } \\
\hline
{$\mu$} &{$<10^{-18}$ } (CERN)\cite{cern3} & {$10^{-35}$ }\\
 & $\sim10^{-19}$ $^\dag$ (E821) \  & \\
\hline
$^{199}Hg$ & $ 2.1 \times  10^{-28}$  (95\%CL)\cite{hgedm} \\
\hline
 $^\dag$Estimated \\

\end{tabular}
\end{table}

The magnetic dipole moment can differ from its Dirac value
($g = 2$) for several
reasons.  Recall that the proton's $g$-value is 5.6 ($a_p =1.79$), 
a manifestation of
its quark-gluon internal structure. 
On the other hand, the leptons appear to have
no internal structure, and the MDMs are thought
to arise from radiative corrections, i.e. from virtual particles
that couple to the lepton.  We would emphasize that
these radiative corrections need not be limited  to
the standard-model particles, but rather the physical values
of the lepton anomalies represent a sum-rule over {\it all} virtual particles
in nature that can couple to the lepton,
or to the photon through vacuum polarization
loops.

\begin{figure}[h!]
\begin{center}
  \includegraphics[width=0.45\textwidth,angle=0]{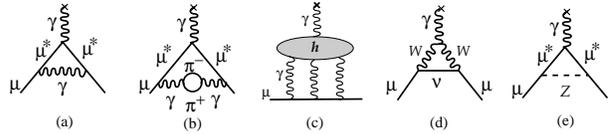}
\end{center}
  \caption{The Feynman graphs for: (a) Lowest-order QED (Schwinger) term;
 (b)Lowest-order hadronic contribution; (c) hadronic light-by-light
contribution;  (d)-(e) the lowest order electroweak $W$ and $Z$ contributions.
The * emphasizes that in the loop the muon is off-shell.  With the present 
limits on $m_h$, the contribution from the single Higgs loop is negligible.

  \label{fg:radcor}}

\end{figure}

The standard model value of a lepton's anomaly,
$a_\ell$, has  contributions from three different sets of radiative
processes: quantum electrodynamics (QED)-- with loops containing 
leptons ($e,\mu,\tau$) and photons; 
hadronic -- with  hadrons in vacuum polarization
 loops;
and weak -- with loops involving the  bosons $W,Z,$ and Higgs.
Examples  are shown in Fig.~\ref{fg:radcor}.  Thus
\begin{equation}
a_{e,\mu}^{{\rm ( SM)}}
= a_{e,\mu}^{({\rm QED})} + a_{e,\mu}^{({\rm hadronic})} +
a_{e,\mu}^{({\rm weak})}\, .
\end{equation}

 The dominant contribution from quantum electrodynamics (QED),
called the Schwinger term\cite{Sch48}, $a^{(2)} = \alpha/2 \pi$,
and is shown diagrammatically in Fig.~\ref{fg:radcor}(a).
The QED contributions have been calculated through four loops, with the
leading five-loop contributions calculated\cite{kno06b}. 
 Examples of the hadronic  and weak contributions
are given in Fig.~\ref{fg:radcor}(b)-(e).

The hadronic contribution
cannot be calculated directly from QCD, since the energy scale
is very low ($m_{\mu} c^2$), although Blum has performed 
a proof of principle calculation on the lattice\cite{blum}.  
Fortunately dispersion theory\cite{miller07} 
gives a relationship between the vacuum polarization loop
and the cross section for $e^+ e^- \rightarrow {\rm hadrons}$,
\begin{equation}
a_{\mu}({\rm Had;1})=({\alpha m_{\mu}\over 3\pi})^2
\int^{\infty} _{4m_{\pi}^2} {ds \over s^2}K(s)R(s);
\end{equation}
$R\equiv  \left\{ {\sigma_{\rm tot}(e^+e^-\to{\rm hadrons})}\right\} /
\left\{ \sigma_{\rm tot}(e^+e^-\to\mu^+\mu^-)\right\}$,
and experimental data are used as input\cite{miller07,hadcom}

The muon anomaly is sensitive to a number of potential candidates 
for physics beyond the standard model\cite{Czar01}:
\begin{enumerate}  
\item muon
substructure, where the contribution depends on the substructure scale
$\Lambda$ as
\begin{equation}
\delta a_{\mu} (\Lambda_{\mu}) \simeq {m^2_{\mu} \over \Lambda^2_{\mu}},
\end{equation}
\item $W$-boson substructure.
\item new particles that couple to the muon,
such as the supersymmetric partners of the weak gauge bosons,
\item extra dimensions
\end{enumerate}

 The potential contribution from
supersymmetry has generated a lot of attention\cite{Martin03,Stockinger_07},
the relevant diagrams are shown in Fig.~\ref{fg:susy} below.  A simple model
with equal masses\cite{Czar01} gives
\begin{equation}
 a_{\mu}^{({\rm SUSY})} 
\simeq {\alpha(M_Z) \over 8 \pi \sin^2 \theta_W} {m^2_{\mu} \over \tilde
    m^2}
\tan \beta \left( 1 - {4\alpha \over \pi}\ln {\tilde m \over m_{\mu}}\right)
\end{equation}
\begin{equation}
\simeq ({\rm sgn} \mu) \ 13 \times 10^{-10}\ \tan \beta\ 
\left({100\ {\rm GeV}  \over \tilde m}\right)^2
\end{equation}
where $\tan \beta$ is the ratio of the two vacuum expectation values of the
two Higgs fields.  If the SUSY mass scale were known, 
then $ a_{\mu}^{({\rm SUSY})} $ would provide a clean way to determine
$\tan \beta$.

\begin{figure}[h!]
\begin{center}
  \includegraphics[width=0.3\textwidth,angle=0]{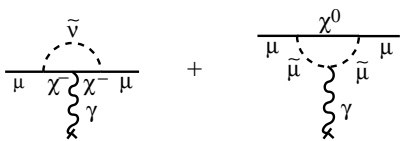}
\end{center}
  \caption{The lowest-order supersymmetric contributions to the muon 
anomaly. The $\chi$ are the superpartners of the standard-model gauge bosons.
  \label{fg:susy}}

\end{figure}

\subsection{Measurement of the Anomalous Magnetic Dipole Moment}

Measurement of the magnetic anomaly uses the spin motion in a magnetic
field.  For a muon moving in a magnetic field, the spin and momentum
rotate  with the frequencies:
\begin{equation}
\vec \omega_S= - {qg \vec B \over 2m}  
- {q \vec B \over \gamma m }(1-\gamma);
\ \ \vec  \omega_C = - {q \vec B \over m \gamma}.
\label{eq:freq-spin-cyc}
\end{equation}
The spin precession {\it relative to the momentum} 
occurs at the difference frequency, $\omega_a$, between the spin 
and cyclotron frequencies, Equation~\ref{eq:freq-spin-cyc},
\begin{equation}
\vec \omega_a = \vec \omega_S - \vec \omega_C 
= - \left( {g-2 \over 2} \right) {q \vec B \over m}
= - a_\mu { q \vec B \over m}.
\label{eq:diffreq}
\end{equation}
 The magnetic field in Eq.~\ref{eq:diffreq}
is the average field seen by the ensemble of muons.
This technique has been used in all but the
first experiments by Garwin, et al.\cite{garwin}, which used
stopping muons, to measure the anomaly.  After Garwin, et al., 
made a 12\% measurement of the anomaly, a series of three beautiful
experiments at CERN culminated with a 7.3~ppm measure of $a_\mu$\cite{cern3}.

In the third CERN experiment, a new technique
was developed based on the observation that electrostatic quadrupoles
could be used for vertical focusing.  With 
the velocity transverse to the magnetic field
($\vec \beta \cdot \vec B = 0$), the spin precession formula becomes
\begin{equation}
\vec \omega_a = 
 - \ {q \over m} 
\left[ a_{\mu} \vec B -
\left( a_{\mu}- {1 \over \gamma^2 - 1} \right)
{ {\vec \beta \times \vec E }\over c }
\right]\,.
\label{eq:omega}
\end{equation}
For $\gamma_{\rm magic} = 29.3$, ($p_{magic} = 3.09$~GeV/$c$), the 
second term vanishes; one is left with the simpler result of
Equation~\ref{eq:diffreq}, hence the name ``magic'', and the electric
field does not contribute to the spin precession relative to the 
momentum. There are two major advantages of using the
magic $\gamma$ and a uniform magnetic field: 
(i) the knowledge needed on the muon trajectories to determine the
average magnetic field is much less than when gradients are present,
and the more uniform field permits NMR techniques to realize their 
full accuracy, thus increasing the knowledge of the $B$-field.
The spin precession is determined almost completely by
Eq.~\ref{eq:diffreq}, which is independent of muon momentum; {\it all}
muons precess at the same rate.
This technique was used also in experiment E821\cite{bennett} at 
the Brookhaven National Laboratory Alternating Gradient Synchrotron (AGS).
The reader is referred to Ref.~[38]
for a discussion of muon decay relevant to E821,
and to Ref.~[40] for details of E821.

\begin{figure}[h]
\includegraphics[width=0.44\textwidth]{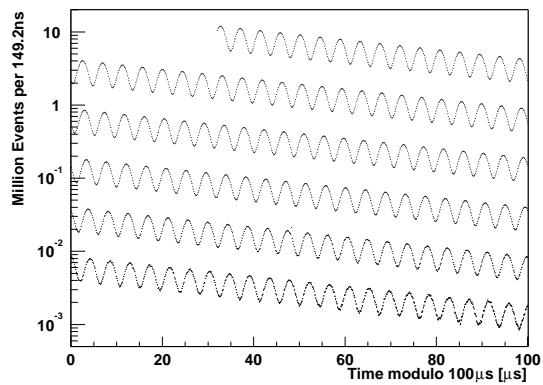}
\caption{The time spectrum of $3.6 \times 10^9$ 
electrons with energy greater than
1.8 GeV from the 2001 data set. 
The diagonal ``wiggles'' are displayed  modulo 100~$\mu$s.
\label{fg:wiggles}}
\end{figure}

Muons are stored in a storage ring\cite{bennett}, and the arrival
time and energy of the decay electrons is measured.
When a single energy threshold is placed on the decay electrons,
the number of high-energy
electrons is modulated by the spin precession frequency, Eq~\ref{eq:omega},
producing the time distribution
\begin{equation}
N(t,E_{th}) = 
N_{0}(E_{th})e^\frac{-t}{\gamma\tau}
[1+A(E_{th})\cos(\omega_{a}t+\phi(E_{th}))].
\label{eq:fivep} 
\end{equation}
as shown in Fig.~\ref{fg:wiggles}\cite{bennett}.
 The value of $\omega_a$ is obtained from a 
least-squares fit to these data.  
The five-parameter
function (Eq.~\ref{eq:fivep}) is used as a starting point, but  
many additional small effects must be taken into 
account\cite{bennett}.

\begin{figure}[h]
\begin{center}
\includegraphics[width=0.3\textwidth,angle=0]{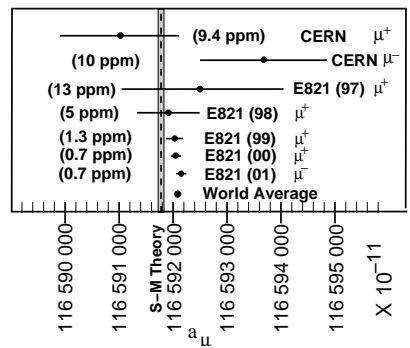}
\caption{Measurements of the muon anomaly, indicating the value, as well as
the muon's sign.  As indicated in the text, to obtain
the value of $a_{\mu^-}$ and the world average
 {\sl CPT} invariance is assumed. The theory value is taken from
Ref.~[38], which uses electron-positron
annihilation to determine the hadronic contribution.
\label{fg:g2results}}
\end{center}
\end{figure}

In E821, both $\mu^+$ and $\mu^-$ were measured, and
assuming {\sl CPT} invariance, the final result obtained by  
E821\cite{bennett}, 
$a_{\mu}^{\mbox{\rm\tiny
exp}}~=116\,592\,080~(63)\times
10^{-11}$,
is shown in 
Fig.~\ref{fg:g2results}, along with the individual measurements and
the standard-model value.
 The present standard-model value is\cite{miller07}
$a_{\mu}^{\mbox{\rm\tiny SM(06)}}=116\,591\,785~(61)\times 10^{-11}$,
and one finds $\Delta a_{\mu}=295(88)\times 10^{-11}$,
a  $3.4~\sigma$ difference.

\begin{figure}[h!]
\begin{center}
  \includegraphics[width=.3\textwidth]{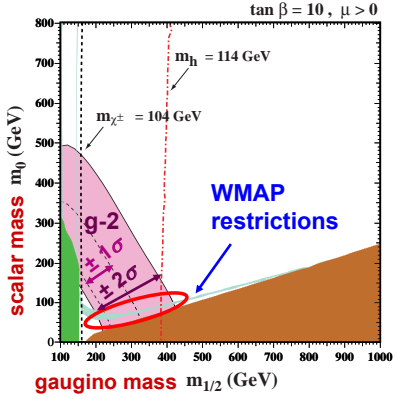}
\end{center}
  \caption{Limits on dark matter placed by various inputs in CMSSM,
with $\tan \beta = 10$.
The $\Delta$ between experiment and standard-model theory is from
Ref.~[38], see text. 
The brown wedge on the lower right is excluded by
the requirement the dark matter be neutral.  Direct limits on
 the Higgs and chargino $\chi^\pm$ masses are indicated by vertical
lines. Restrictions from the WMAP satellite data are shown as a 
light-blue line. The $(g-2)$ 1 and 2-standard deviation boundaries are shown
in purple.
The region ``allowed'' by WMAP and $(g-2)$  is indicated by the ellipse,
which is further restricted by the limit on $M_h$.
 (Figure courtesy of K. Olive)
}  
\label{fg:dark}
\end{figure}

One candidate for the cosmic dark matter is the lightest
supersymmetric partner, the neutralino, $\chi^0$ in Fig.~\ref{fg:susy}. 
In the context of a constrained minimal
supersymmetric model (CMSSM), $(g-2)_\mu$ provides an orthogonal
constraint on dark matter\cite{olive} from that provided by the 
WMAP survey, as can be seen in
Fig.~\ref{fg:dark}.

With the apparent 3.4 $\sigma$ difference between theory and experiment,
a new experiment to improve the error by a factor of 2 to 2.5 has been
proposed to 
Brookhaven laboratory, but at present it is not funded.  
The theoretical value
will continue to be improved, both with the expected availability of 
additional data on $e^+e^-$ annihilation to hadrons, and with additional 
work on the hadronic light-by-light contribution\cite{miller07}.

\subsection{The Search for a Muon Electric Dipole Moment}

With an EDM present, the spin precession frequency
relative to the momentum must be  modified.
The total frequency becomes
 $\vec \omega = \vec \omega_a + \vec \omega_\eta$, where
\begin{equation}
\vec \omega_\eta =
- \frac {q}{m}\left[ {\eta \over 2} \left( {\vec E \over c} +
\vec \beta \times \vec B \right) \right] \, ,
\label{eq:omegaeta}
\end{equation}
with $\eta$ defined by Eq.~\ref{eq:eta}, and
$\omega_a$ by Eq.~\ref{eq:omega}.
The spin motion resulting from the motional electric 
field, $\vec \beta \times \vec B$ is the dominant effect, so
$\omega_\eta$ is transverse to $\vec B$.  An EDM would 
have two effects on the precession, 
there would  be a slight 
tipping of the precession plane, which would cause a vertical oscillation of
the centroid of the decay electrons that out of phase with the
$\omega_a$ precession; and  the observed frequency $\omega$
would be larger,
\begin{equation}
\omega=\sqrt{\omega_a^2+ \left({q\eta\beta B\over 2m}\right)^2}\, .
\label{eq:omEDM}
\end{equation}
 The muon limit in Table~\ref{tb:edm}, 
placed by the non-observation of the vertical oscillation, is dominated
by systematic effects.
The limit obtained by this method in the CERN experiment,\cite{cern3}
and likely to be obtained by E821, cannot directly exclude the possibility
that the entire difference between the measured and standard-model
values of $a_\mu$ could be caused by a
muon EDM.  Such a scenario would imply that the EDM would be
$d_{\mu}=2.4(0.4)\times 10^{-19}$ e-cm, a
 factor $\approx 10^8$ larger than the current limit on the
electron EDM.  While this would be a very exciting result, it is
orders of magnitude larger than that expected from even the most
speculative models\cite{ellis1,ellis2,bdm,feng}

To reduce systematic errors in the muon EDM measurement,
a ``frozen spin'' technique has been proposed\cite{farley04} which uses
a radial electric field in a muon storage ring, operating at
 $\gamma <<\gamma_{\rm magic}$
 to cancel the $(g-2)$ precession.  The EDM term, Eq.~\ref{eq:omegaeta},
would then cause the spin to steadily move out of the plane of the storage
ring. Electron detectors above and below the storage region would detect a
time-dependent up-down asymmetry that increased with time.  As in the $(g-2)$
experiments,  detectors placed in the plane of the beam would be used,
in this case to make sure that the radial-$E$-field cancels the normal
spin precession exactly. Adelmann and Kirsh\cite{AK} have proposed
 that one could reach a 
sensitivity of $5 \times 10^{-23}\,e-$cm with a small storage ring at PSI.
A letter of intent at J-PARC\cite{loi} suggested that one could reach 
$< 10^{-24}\,e-$cm there.  The ultimate sensitivity would need 
an even more intense muon source, such as a neutrino factory.

\section{The Search for Lepton Flavor Violation}

The standard-model gauge bosons do not permit leptons to mix with 
each other, unlike the quark sector where
mixing has been known for many years.  Quark mixing
was  first proposed by 
Cabibbo\cite{cabibbo}, and extended to three generations by
Kobayashi and Maskawa\cite{KM}, which is described
by a mixing $3\times 3$ matrix  now universally called the  CKM matrix.
With the discovery of neutrino mass, we know that
lepton flavor violation (LFV) certainly exists in the 
neutral lepton sector, with the determination of the 
mixing matrix for the three neutrino
flavors having become a 
world-wide effort. 

\begin{figure}[htb]
\centerline{\includegraphics[width=0.4\textwidth]{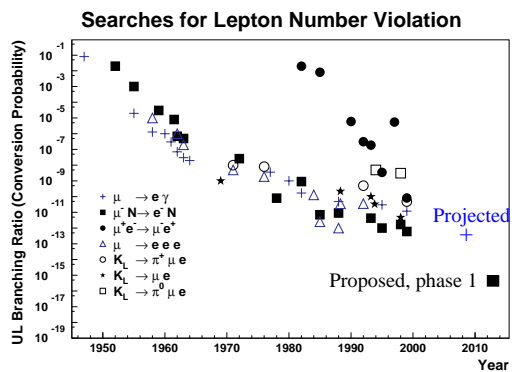}}
 \caption{Historical development of the 90\% C.L. upper limits
  on branching
ratios
 respectively conversion probabilities
 of muon-number violating processes which involve muons and kaons.
Also shown is the projected goal of the MEG ($\mu^+ \rightarrow e^+ \gamma$)
 experiment which is underway at PSI, and the projected sensitivity of
the recent letter of intent to J-PARC for muon-electron conversion.
(Figure from Ref.~[13])}
 \label{fg:history}
 \end{figure}

While the mixing observed in neutrinos does predict some level of charged
lepton mixing, it is many orders of magnitude below present experimental 
limits\cite{cernwg}.  
New 
dynamics\cite{Barbieri95,Hasegawa,kitano1,kitano2,Barenboim,
Gouvea,Cirigliano,Borzumati,
masiero,paradisi}, 
e.g. supersymmetry, do permit leptons to mix, and
the observation of standard-model forbidden processes such as
\begin{equation}
\mu^+  \rightarrow e^+ \gamma;  \quad  \mu^+  \rightarrow e^+ e^+ e^-;
 \quad \mu^- N  \rightarrow e^- N;
\end{equation}
\begin{equation}
 \mu^+e^-  \rightarrow \mu^- e^+; \quad 
\mu^- + {\mathcal N} \rightarrow e^+ + {\mathcal N'}
\label{eq:delta2}
\end{equation}
would clearly signify the presence of new physics. 
 The present
limits on lepton flavor violation are shown in Fig.~\ref{fg:history}.

\begin{figure}[h!]
\begin{center}
  \includegraphics[width=.45\textwidth]{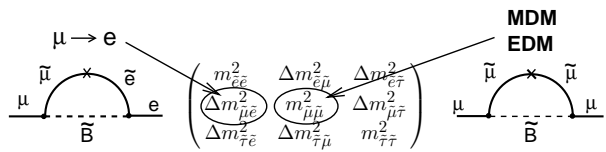}
\end{center}
  \caption{The supersymmetric contributions to the anomaly, and to
$\mu \rightarrow e$ conversion, showing the relevant slepton mixing matrix
elements. The MDM and EDM give the real and imaginary parts of
the matrix element respectively.
\label{fg:susymix}}
\end{figure}

If lepton mixing occurs via  supersymmetry, there will be a mixing between
the supersymmetric leptons (sleptons) which would also be described by a
$3 \times 3$ mixing matrix.  The schematic connection between lepton flavor
violations and the dipole moments is shown in Fig.~\ref{fg:susymix},
and there are models that try to connect these processes\cite{baek}.

In a large class of models, if the $\Delta \ell = 1$ {\sl LFV} 
decay goes through the transition magnetic moment, one finds\cite{cernwg}
\begin{equation}
{ {B(\mu N \rightarrow e N)} \over {B(\mu \rightarrow e \gamma)}} =
2\times 10^{-3} B(A,Z),
\end{equation}
where $B(A,Z)$ is a coefficient of order 1 for nuclei heavier than 
aluminum\cite{cmm}.  For other models, these two rates can be
the same\cite{cernwg}, so in the design of new experiments, the reach
in single event sensitivity for the coherent muon conversion experiments
needs to be several orders of magnitude smaller
than for $\mu\rightarrow e \gamma$ to probe the former class of models
with equal sensitivity. Detailed calculations 
of $\mu-e$ conversion rates as a function of atomic
number have also been carried out\cite{kitano3}, and
if observed, measurements should be carried out in several nuclei.

\begin{figure}[h!]
\begin{center}
\includegraphics*[width=.35\textwidth]{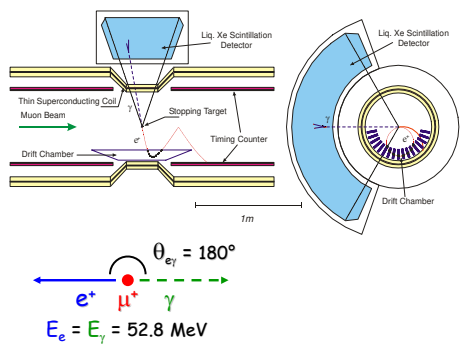}
\end{center}
\caption{The side and end views of the 
MEG experiment. Since the muon is at rest,
the photon and positron are at a relative angle of
 $180^{\circ}$.  The positron is
tracked in a magnetic field which produces a 
constant bending radius, independent
of angle. }
\label{fg:meg}
\end{figure}

From the experimental side, 
 the next generation
$\mu \rightarrow e \gamma $ experiment, MEG, is now under
way at PSI\cite{meg}, with a sensitivity goal of $10^{-13} - 10^{-14}$. 
 Since the decay occurs at rest, the photon and positron are
 back-to-back, and share equally
the energy $m_\mu c^2$.  This experiment makes 
use of a unique  ``COBRA'' magnet which produces a constant bending radius
for the mono-energetic $e^+$, independent of its angle.  The photon is detected
 by a large liquid Xe scintillation detector as shown in Fig. \ref{fg:meg}.

Of the various lepton-flavor violating reactions,
only coherent muon conversion does not require coincidence measurements.
The decay $\mu \rightarrow  3e $, while theoretically appealing,
requires a triple coincidence and sensitivity to the whole phase space of
the decay, and thereby is experimentally more challenging.  It is 
the coherent muon to electron conversion, where with adequate energy
resolution, the conversion electron can be resolved from background,
that with adequate muon flux can be pushed to the $10^{-18}$ or $10^{-19}$
sensitivity.  Such a program has been proposed for J-PARC\cite{prism-prime}.

The muonium to antimuonium conversion 
(left-hand process in Eq.\ref{eq:delta2})
represents a change of 
two units of lepton number, analogous to $K^0$ $\bar K^0$ oscillations.
This process was originally proposed by Pontecorvo\cite{pont}.  An
experiment at PSI\cite{willmann} obtained a single event sensitivity of
$P_{M \bar M} = 8.2 \times 10^{-11}$ which implies a coupling
$G_{M \bar M }\leq 3 \times 10^{-3} G_F$ at 90\%\ C.L., where
$G_F$ is the Fermi coupling constant.
A broad range of speculative theories such as left-right
symmetry, R-parity violating supersymmetry, etc.\cite{mmb_theories},
could permit such an oscillation.

\section{Summary and Conclusions}

Since its discovery, the muon has provided an important tool to study the
standard model, and to constrain its extensions.  Experiments in the planning
stage for $(g-2)$, the search for an electric dipole moment and
lepton flavor violation in muon decay or conversion will continue this 
tradition. Research and development for new more intense muon sources,
such as the muon ionization cooling experiment (MICE)\cite{mice},
will further propel increases in sensitivity. 
Muon experiments form an important part of the precision 
frontier in particle physics, 
which will continue to provide vital information complementary to that from
the highest energy colliders.

I wish to acknowledge A. Czarnecki, T. Gorringe, D. Hertzog,
P. Kammel,  K. Jungmann, Y. Kuno,
W. Marciano, J.P. Miller, Y. Okada, and E. de Rafael,  
for helpful conversations,
and TG, DH, KJ,
PK and JM for their excellent suggestions on this manuscript.  This work was 
supported in part by the U.S. National Science Foundation and U.S. Department
of Energy.


\begin{thebibliography}{9}
\bibitem{convention}The 
standard notation in the literature for the hadronic current
has changed three times over the years.  I follow the notation of Gorringe and 
Fearing in Ref.~[18].
\bibitem{hadcom} It has been proposed that the hadronic contributions 
could also be determined from hadronic $\tau$ decay, plus the conserved
vector current hypothesis, but this prescriptions seems to have
internal consistency issues which are still under study.  See
Ref.~[37] and references therein.
\end{thebibliography}
\end{document}